\documentclass[runningheads]{llncs}
\usepackage{graphicx}
\usepackage[T1]{fontenc}
\usepackage{changepage}
\usepackage{amsfonts}
\usepackage{bbding}

\usepackage[whole]{bxcjkjatype} 
\usepackage{float}
\usepackage{enumitem}

\begin{document}

\title{Remote Verification System for Mizar Integrated with Emwiki}

\author{
    Toshiki Kai\inst{1} \and
    Yuta Teruya\inst{1} \and
    Kazuhisa Nakasho\inst{1}\Envelope\orcidID{0000-0003-1110-4342}
}
\authorrunning{T. Kai et al.}
\institute{
    \bigskip 
    \textsuperscript{1} Yamaguchi University, 2-16-1 Tokiwa-dai, Ube, Japan \\
    \email{nakasho@yamaguchi-u.ac.jp} \\
    \url{https://www.eng.yamaguchi-u.ac.jp/}
}
\maketitle

\begin{abstract}
In this paper, we present a remote verification environment for Mizar and its integration with a web platform. Although a VSCode extension for Mizar is already available, it requires installing the Mizar verification tools locally. Our newly developed system implements these verification environments on a server, eliminating this requirement. First, we explain the implementation of the remote verification environment for Mizar and the VSCode for the Web extension. Second, we discuss the integration with the web platform emwiki, which allows browsing the existing Mizar Mathematical Library (MML).

\keywords{Mizar \and Remote Development \and Web Platform Integration \and VSCode for the Web Extension \and Mizar Mathematical Library (MML).}

\end{abstract}
\section{Introduction}
Currently, Mizar Extension\footnote{\url{https://marketplace.visualstudio.com/items?itemName=fpsbpkm.mizar-extension}}~\cite{taniguchi2021visual} and Mizar Mode~\cite{bancerek2004integrated} are provided as editor environments for Mizar~\cite{grabowski2010mizar,bancerek2015mizar}, an interactive theorem prover (ITP). To use these editor environments, users need to install the Mizar system and editors in their local environment. In this study, we developed a remote verification environment and a VSCode for the Web extension to improve the user experience for Mizar users and realized a Mizar integrated development environment (IDE) that does not require a local setup.
Furthermore, the remote development environment is linked to the emwiki system, a documentation tool for the Mizar Mathematical Library (MML), to enhance the usability of the development environment. The previous emwiki system was primarily used for searching and browsing libraries. Although it had a function for adding comments to Mizar source code, it lacked the capability to develop new libraries. With the addition of this new feature, the emwiki system has evolved into a comprehensive web-based authoring tool for the Mizar library.


\section{Related Works}
\subsection{Remote Verification Environements for ITPs}
\textbf{Lean 4 Web}\footnote{\url{https://lean.math.hhu.de/}} is a web-based integrated development environment (IDE) for Lean 4 that runs the Lean server on a web server. It offers features such as document browsing with hover-based information display and real-time feedback with error checking, enabling efficient development in the Lean 4 language.

\textbf{jsCoq}\footnote{\url{https://jscoq.github.io/}} is a system that runs Coq in a web browser. JsCoq includes IDE functions for efficiently editing Coq scripts and embedding them as HTML documents, allowing users to check the progress of proofs in real-time~\cite{arias2017jscoq}.

\textbf{Clide}\footnote{\url{https://github.com/martinring/clide2}} is a web interface for Isabelle that extends the CodeMirror editor and is built utilizing the Isabelle/PIDE framework~\cite{wenzel2018isabelle}. With Clide, proofs can be edited collaboratively, and changes are synchronized to all users in real-time~\cite{luth2013web,Ring2014Clide}.

\textbf{Isabelle/Cloud}~\cite{xu2023isabelle}, by Hao Xu and Yongwang Zhao, delivers Isabelle/HOL as a cloud-based IDE using VSCode for the Web. This setup promotes efficient theorem proving without local installations. Our work extends this approach to support Mizar, enhancing accessibility and collaborative formal verification.

\subsection{Existing Utilities for Mizar}
\textbf{Bancerek's Remote Verifier} was the first Mizar online verifier. It was developed by Grzegorz Bancerek and published on the Mizar forum in 1997\footnote{\url{https://mizar.uwb.edu.pl/forum/archive/9708/msg00005.html}}.
That old service is no longer available, but you can find its successor, which is still usable\footnote{\url{http://fm.uwb.edu.pl/classes/anonymous.php?page=/remote/}}.

\textbf{MizAR}~\cite{kaliszyk2015mizar,jakubuuv2023mizar}, a much more modern online Mizar verifier developed by Josef Urban, integrates machine learning with automated reasoning to enhance the Mizar system. It uses models trained on the MML to predict useful lemmas and theorems, improving the success rate of verification. By increasing proof automation, MizAR aids mathematicians in verifying complex theorems more efficiently. This system is also available on the Web\footnote{\url{http://grid01.ciirc.cvut.cz/~mptp/MizAR.html}}. 

\textbf{Mizar Extension}~\cite{taniguchi2021visual} is an extension available for Visual Studio Code. It provides features such as syntax highlighting, code formatting, execution and halting of Mizar commands, definition jumps, and hover functions for referencing cited theorems and definitions. Additionally, Mizar Extension displays progress information when executing Mizar commands.

\textbf{The emwiki system}\footnote{\url{https://em1.cs.shinshu-u.ac.jp/emwiki/release/}} is a web platform for hosting the Mizar Mathematical Library (MML)~\cite{yamamichi2022emwiki,furushima2022integrated}. Developed as a successor to the MathWiki Project~\cite{alama2011large,tankink2013formal}, emwiki aims to improve the readability and accessibility of Mizar articles. The platform's modular architecture ensures consistency, extensibility, and interoperability. Its main features include:

\begin{enumerate}[leftmargin=*,topsep=0ex]
\item Wiki function: enables users to add comments to the HTMLized MML~\cite{furushima2022integrated}.
\item Search function: allows users to search for articles, symbols, and theorems in the MML~\cite{nakasho2015documentation}.
\item Graph display function: visualizes the dependency relationships between MML articles~\cite{shigenaka2022emgraph,suzuki2023classification}.
\end{enumerate}

\section{Remote Development Environment}
\subsection{Mizar Verification Server}
The remote verification environment developed in this study requires users to access the Mizar verification program from a web browser. However, the Mizar verification program does not support WebAssembly. To address this, we developed a new web application called \textit{Mizar Server} that executes Mizar commands on a remote server. This eliminates the need for Mizar users to maintain a local development environment or to update the Mizar Mathematical Library (MML). Users of the system can access the various functions of Mizar Server via HTTP requests. 

\begin{figure}[htbp]
    \centering
    \includegraphics[width=1.0\linewidth,clip]{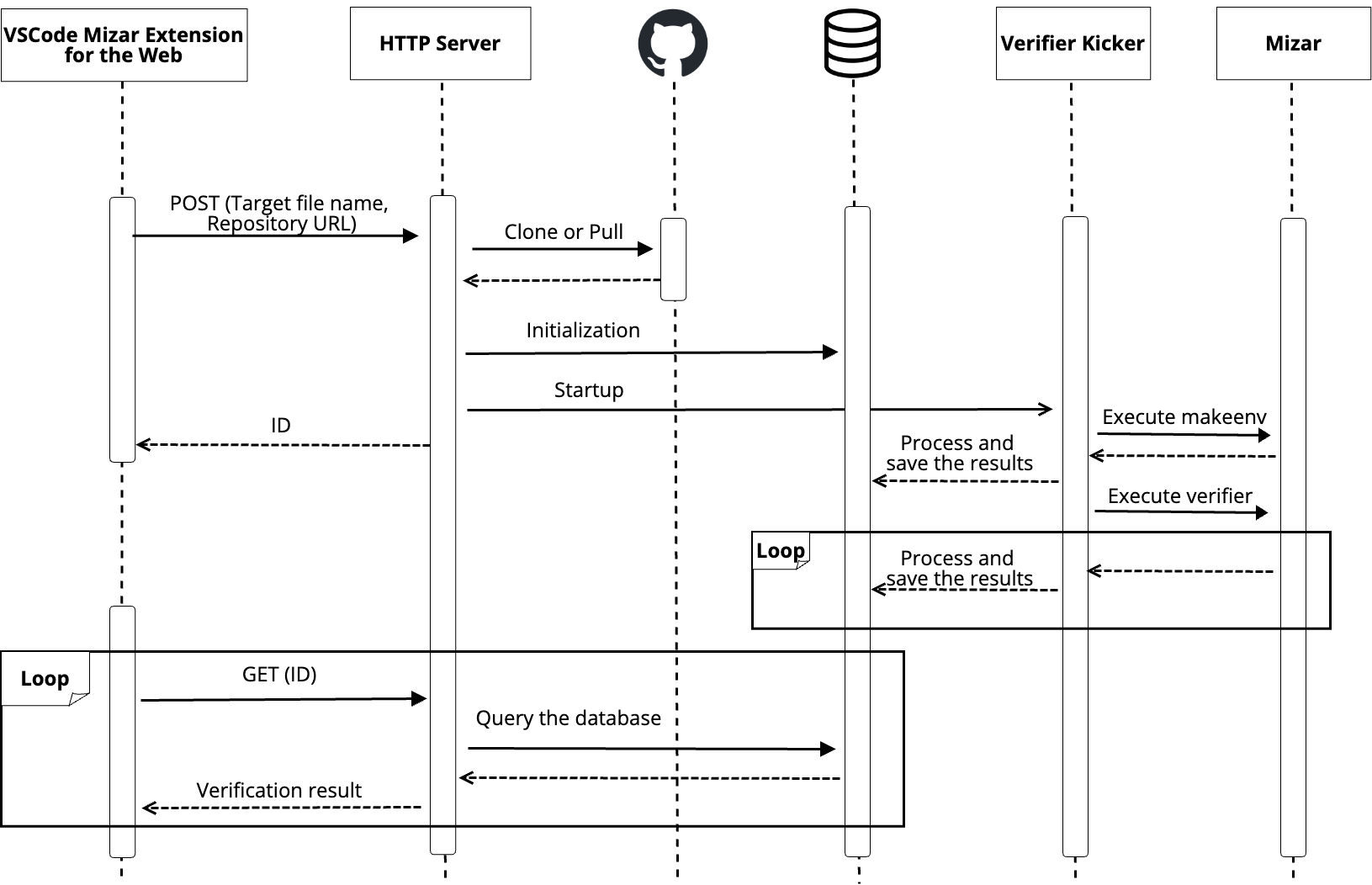}
    \caption{Sequence diagram of verifier execution}
    \label{fig:sequence_diagram}
\end{figure}

When Mizar Server receives a request, it executes a command program such as verifier and immediately returns the ID associated with the request as a response. Mizar Server provides an API that periodically requests the verification progress using the ID attached to the request as a parameter. This API informs the user about the progress rate if the verifier or other commands are still running, and notifies the user of any errors once the command execution is completed. The sequence diagram of the verifier command execution is shown in Fig. \ref{fig:sequence_diagram}.

Mizar Server also provides source code linter and formatter. Additionally, by specifying the Mizar version during verification, it is possible to execute Mizar binaries of different versions. All APIs provided by Mizar Server use TLS/SSL communication, ensuring secure communication. This prevents the interception and tampering of communication data on the network and protects users' confidential information.

\subsection{VSCode for the Web Extension}
VSCode for the Web\footnote{\url{https://vscode.dev/}} is a cloud-based code editor developed by Microsoft. Users can create projects, edit code, and manage versions via a web browser. By making VSCode's functionality available on a web browser, it reduces the effort required for installation and configuration, providing a consistent development environment across devices. We have ported all the functionality of the Mizar Extension to the VSCode for the Web extension.

\begin{figure}[htbp]
    \centering
    \includegraphics[width=\linewidth]{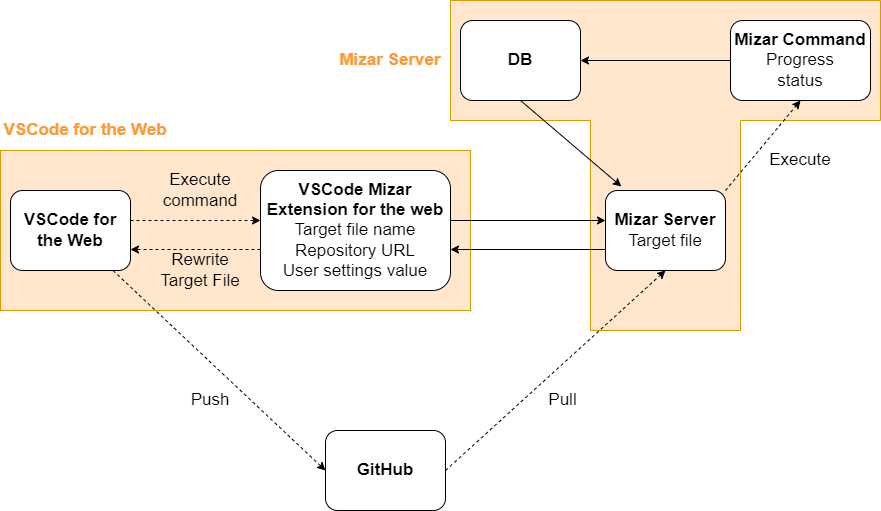}
    \caption{Architecture diagram of Mizar Server and VSCode for the Web}
    \label{fig:architecture_diagram}
\end{figure}

Mizar Server and VSCode for the Web share Mizar source code via GitHub to ensure that the Mizar source code in the development environment is synchronized. The architecture diagram of Mizar Server and VSCode for the Web is shown in Fig. \ref{fig:architecture_diagram}.  This work builds on our preliminary version presented in \cite{matumoto2023vscode}, and Fig. \ref{fig:sequence_diagram} and Fig. 
\ref{fig:architecture_diagram} are adapted from that earlier version.

The current versions of Mizar Server and VSCode for the Web Extension that we developed do not support the Language Server Protocol (LSP). This limitation exists because the existing VSCode Extension for Mizar~\cite{taniguchi2021visual} did not support LSP, and our development extends this existing extension. In the future, when supporting editors other than VSCode for the Web, we plan to build a server that complies with LSP, as many other ITP language servers have adopted.

\subsection{Linking emwiki and Remote Verification Environment}

By integrating the functionality of the Mizar remote verification environment into emwiki, users can now browse the library, create explanatory articles, and develop MML on a single platform. This allows users to seamlessly proceed with Mizar development and documentation. 
The goal of this system is to provide a useful IDE for novice users who are not familiar with ITPs. However, the setup procedure in the current version is somewhat complicated. We assume that users of our system are familiar with GitHub, as the setup includes operations such as forking the repository and preparing a classic OAuth Token. Detailed instructions for setting up the environment can be found by logging into emwiki and opening the Develop tab\footnote{\url{https://em1.cs.shinshu-u.ac.jp/emwiki/release/settings/develop}}.



\subsection{Implemented Features}
\begin{figure}[htbp]
    \centering
    \includegraphics[width=\linewidth]{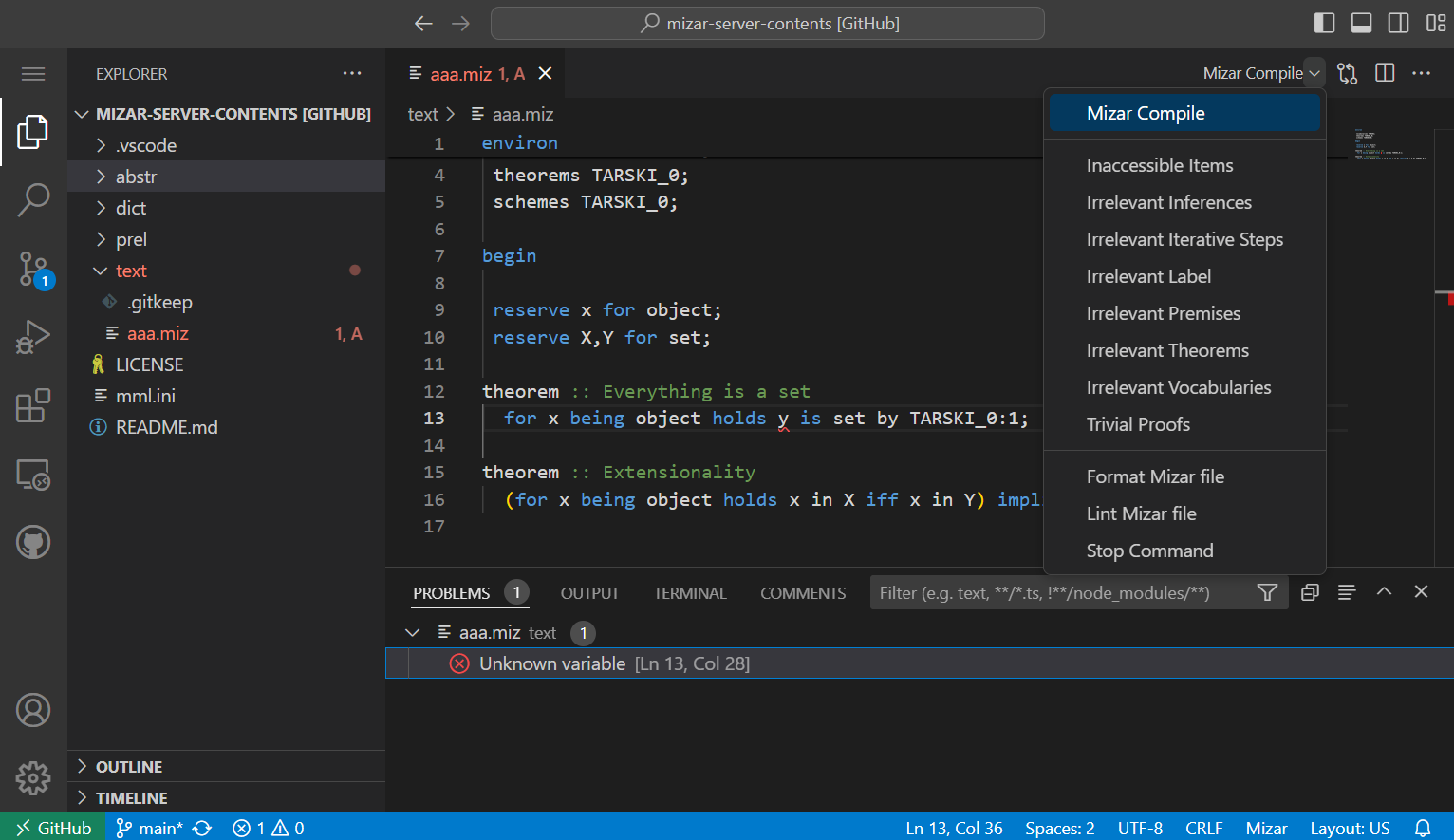}
    \caption{Screenshot of VSCode for the Web extension}
    \label{fig:ide_screenshot}
\end{figure}

Our VSCode for the Web extension has ported all the functions implemented in the existing Mizar Extension~\cite{taniguchi2021visual}, including syntax highlighting, code formatting, execution and halting of Mizar commands, definition jumps, and hover functions. After setting up the environment, the user creates a `.miz' file in the ``text'' folder to write a Mizar article. To verify the proof, invoke the command palette by pressing [CTRL+P] while the cursor is in the editor and select ``Mizar: Mizar Compile,'' or choose ``Mizar Compile'' from the command selection menu located in the upper right corner on the editor. Commands for Mizar brush-up tools, Formatter and Linter are executed in the same way. A command cancel function and a progress bar are also implemented. If errors are found during verification, they are highlighted with wavy lines after the verification command is executed, and the errors are displayed in the PROBLEMS window. Fig. \ref{fig:ide_screenshot} shows a screenshot of the editor.

\section{Evaluation}
The Mizar remote verification environment developed in this study frees Mizar library developers from the responsibility of setting up a local Mizar development environment. However, remote verification faces the challenge of reduced execution speed of the verification function due to the overhead of exchanging Mizar files over the network. We compared the performance of Mizar verification using the existing VSCode Extension and our remote verification environment on a Mizar file with 3,657 lines of code. The results showed that the local verification environment took 10.02 seconds, while the remote verification environment took 12.27 seconds, indicating that the remote environment was about 2 seconds slower. This overhead arises from the fact that the source code is sent to the Mizar server via GitHub. Although using GitHub facilitates source code synchronization and management, it increases the communication overhead.

\section{Conclusion and Future Work}
In this paper, we proposed a remote verification environment for Mizar and its integration with emwiki. The application we have built is available at the following URL. To use this feature, users must log in to emwiki.
\begin{description}
    \item [Remote IDE]: \url{https://em1.cs.shinshu-u.ac.jp/emwiki/release/settings/develop}
\end{description}
The source code is available in the following repositories.
\begin{description}
    \item [emwiki]: \url{https://github.com/mimosa-project/emwiki}
    \item [emvscode-web]: \url{https://github.com/mimosa-project/emvscode-web}
    \item [mizar-server]: \url{https://github.com/mimosa-project/mizar-server}
\end{description}
A future challenge is to improve the performance of the remote verification environment. We believe that the parallelization of the verification program and deploying a high-performance verification server will reduce the execution time compared to a typical local environment. Our system avoids local installation but needs emwiki, GitHub and VSCode knowledge. These can make it harder for educational use and beginners. We would like to work on simplifying account management and reducing external dependencies to improve accessibility.

\section*{Acknowledgments}
This work was supported by JSPS KAKENHI Grant Number 24K14897．

\newpage
\bibliographystyle{ieeetr}
\bibliography{reference}
\end{document}